\documentclass[aip,reprint,jcp]{revtex4-1}
\usepackage{color}
\usepackage{graphicx}
\usepackage{dcolumn}
\usepackage{epstopdf}
\usepackage{hyperref}
\usepackage{amsmath}
\usepackage{mhchem}
\usepackage{siunitx}
\DeclareSIUnit\atm{atm}
\DeclareSIUnit\cal{cal}
\DeclareSIUnit\e{e}
\DeclareSIUnit\Debye{D}
\DeclareSIUnit\debye{D}
\DeclareSIUnit\kcalmol{\kilo\cal\per\mol}
\DeclareSIUnit\molal{\mol\per\kilo\gram}
\DeclareSIUnit[number-unit-product = {}]\molar{M}

\newcommand{\s}{$S_{\text{s}}$\,}
\newcommand{\scv}{$S_{\text{CV}}$\,}

\newcommand{\kb}{\,$\text{k}_\text{B}$\,}
\newcommand{\red}[1]{\textcolor{red}{#1}}

\begin{document}

\author{S. P\'erez-Conesa}
\affiliation{Department of Physical Chemistry, University of Seville, 41012 Seville, Seville, Spain }
\author{Pablo M. Piaggi}
\affiliation{Theory and Simulation of Materials (THEOS), {\'E}cole Polytechnique F{\'e}d{\'e}rale de Lausanne 
(EPFL),  CH-1015  Lausanne,  Switzerland}
\affiliation{Facolt{\`a} di Informatica, Istituto di Scienze Computazionali, and National Center for 
Computational Design and Discovery of Novel Materials (MARVEL), Universit{\`a} della Svizzera italiana (USI), 
Via Giuseppe Buffi 13, CH-6900, Lugano, Switzerland}
\author{Michele Parrinello}%
\email{parrinello@phys.chem.ethz.ch}
\affiliation{Department of Chemistry and Applied Biosciences, ETH Zurich, c/o USI Campus, Via Giuseppe Buffi 
13, CH-6900, Lugano, Switzerland}
\affiliation{Facolt{\`a} di Informatica, Istituto di Scienze Computazionali, and National Center for 
Computational Design and Discovery of Novel Materials (MARVEL), Universit{\`a} della Svizzera italiana (USI), 
Via Giuseppe Buffi 13, CH-6900, Lugano, Switzerland}

\title{A local fingerprint for hydrophobicity and hydrophilicity: from methane to peptides}

\keywords{metadynamics, enhanced sampling, 
hydrophobicity,hydrophilicity,solvation,entropy,amino acids}

\begin{abstract}
An important characteristic that determines the behavior of a solute in water is whether it is hydrophobic or hydrophilic.
The traditional classification is based on chemical experience and heuristics. 
However, this does not reveal how the local environment modulates this important property. 
We present a local fingerprint for hydrophobicity and hydrophilicity inspired by the two body contribution to the entropy.
This fingerprint is an inexpensive, quantitative and physically meaningful way of studying hydrophilicity and hydrophobicity that only requires as input the water-solute radial distribution functions. 
We apply our fingerprint to octanol, benzene and the 20 proteinogenic amino acids. 
Our measure of hydrophilicity is coherent with chemical experience and, moreover, it also shows how the character of an atom can change as its environment is changed. 
Lastly, we use the fingerprint as a collective variable in a funnel metadynamics simulation of a 
host-guest system. 
The fingerprint serves as a desolvation collective variable that enhances transitions between the bound and unbound states. 
\end{abstract}

\maketitle

\section{Introduction}
Like dissolves like is one of the earliest chemical rules a scientist learns in relation to solvation. 
It implies that solutes that are chemically similar to water have a favorable interaction with water and are hydrophilic. 
On the other hand, solutes that are not like water will tend to repel water and be hydrophobic\cite{Chandler2005,KaLum1999}. 
Typically one assigns to each atom its own hydrophobicity or hydrophilicity based on chemical experience and heuristics. 
Despite the importance of these intuitive classifications, none of them is quantitative, nor takes into account thermodynamics or solvent structure. 
Processes like protein folding, the assembly of molecules, or crystallization depend crucially on their interaction with water.
Thus it would be of great help to have a measure of the hydrophobicity and hydrophilicity of the atoms in a solute molecule and understand how these parameters change as the environment changes.

In the context of protein science, many hydrophobicity scales for amino acids have been proposed based on empirical or computational data without any definitive consensus\cite{Simm2016}. 
Scales that focus on the hydrophobicity of selected heavy atoms have also been proposed. 
Some of them are based on local compressibility or density fluctuations of the hydration layers of  
 proteins and surfaces\cite{Rossky2010,Xi2017}. \red{There is a vast and at times controversial 
literature on the concept of hydrophyllicity and we do not want to enter into this arena, nor we 
want to replace what is already available in the literature. We introduce a local fingerprint that 
correlates with the commonly accepted notion of hydrophilicity (See Figure S1 of the Supporting 
Information) and can be used cum grano salis as an 
useful indicator. An advantage of our fingerprint is that it can be experimentally measured. }

Here we propose to use a  concept related to density fluctuations, namely the radial 
distribution function (RDF).
Thus we define a local fingerprint that is a function of the RDF between solute atoms and water 
oxygens. 
This fingerprint has been inspired by our previous work on using approximated expressions for the entropy in order to distinguish between solid-like and liquid-like environments\cite{Piaggi2017,Piaggi2017b}. 
We emphasize that the goal of this work is not to calculate the entropy and that we ignore angular correlations that play an important role in a complex liquid such as water
\cite{lazaridis1992entropy,lazaridis1994simulation,lazaridis2000solvent,kinoshita2006pair,
liu2015order,lynden1997hydrophobic,bergman1999topological,lazaridis1998inhomogeneous,Nguyen2012,
li2012computing,Abel2008,Huggins2013,Nguyen2014}.
With respect to other hydrophobicity measures, our fingerprint has the advantage of being easy to compute and to be defined for each atom.
One can thus assess the hydrophobicity of each individual atom and the modifications that result from changes in its environment.
We apply this fingerprint to water and methane as representatives of optimal hydrophilicity and 
hydrophobicity, and to more complex systems such as octanol,
benzene and the 20 proteinogenic amino acids. 

The local fingerprint does not only provide an inexpensive and quantitative assesment of 
hydrophobicity but it is also a suitable collective variable (CV) to describe solvation in enhanced 
sampling simulations.
In many cases solvation and dissolvation represent a kinetic bottleneck in spite of not being the main processes under study.
This is the case in protein folding, ligand binding and crystallization.
We illustrate the usefulness of the fingerprint in enhanced sampling simulations by using it in a funnel metadynamics simulation of a host-guest system.
The fingerprint enhances the transition between the bound and unbound states through a dynamical description of solvation.

\section{Fingerprint for hydrophobicity and hydrophilicity}\label{Theory}
Theory provides an expansion of the entropy of a liquid as a sum of many-body correlation functions\cite{Green52,nettleton1958expression,baranyai1989direct}. 
Inspired by this theoretical framework we propose the following term of the expansion as a local fingerprint for hydrophobicity and hydrophilicity of atom $i$,
\begin{align}
S_{\text{s}}^{i}=-2\pi \rho_{\text{w,loc}} \int_{0}^{\infty}\left\{\right.& g_{i\text{w}
}(r)\ln\left[g_{i\text{w}}(r)\right] \nonumber \\
&-g_{i\text{w}} (r)+1\left.\right\}r^2dr
\label{sij}
\end{align}
where $\rho_\text{w,loc}$ is the local number density of water, and $g_{i\text{w}}(r)$ is the radial distribution function of atom $i$ of the solute and water oxygen atoms. 
The reader should bear in mind that this is not an expression for the excess entropy of the system but rather one of its contributions. 
Calculating the entropy requires including higher order terms and angular correlations\cite{lazaridis1992entropy,lazaridis1994simulation,lazaridis2000solvent,kinoshita2006pair,liu2015order} at a much higher computational cost. 
This defeats our purpose of having an inexpensive, semiquantitative local fingerprint useful also in 
enhanced sampling simulations.
Furthermore, Equation \ref{sij} can be seen in a different light if it is interpreted as a Bregman divergence between $g_{i\text{w}}(r)$ and the perfect gas RDF, i.e. $g(r)=1 \: \forall \: r$ \cite{Piaggi2018}. 
From this point of view it represents a distance between these two functions. 
Equation \ref{sij} is also connected to the Kirkwood-Buff\cite{Kirkwood51} integrals since both are integrals involving the radial distribution function.

It is instructive to calculate the local fingerprint value in the simple case of a spherical cavity 
of radius $R$ embedded in an ideal solvent.
In this particular case the $g_{i\text{w}}(r)$ in equation \eqref{sij} is:
\begin{equation}
  g_{i\text{w}}(r) = 
     \begin{cases}
       0 &\quad \text{if } r \le R \\
       1 &\quad \text{if } r > R
     \end{cases}
\end{equation}
If we introduce this step function in equation \eqref{sij}, the following formula for the local 
fingerprint of a 
cavity of volume $V=\frac{4}{3}\pi R^3$ is obtained:
\begin{equation}
S^\text{cav}_i=-\frac{\rho k_B V }{2}
\label{fingerprint}
\end{equation}
This expression is the leading term of the solvation entropy in the information theory model of hydrophobic 
interactions\cite{garde1996origin,hummer1996information,Chandler2005} if one assumes that the solvent behaves 
ideally. 

\section{Computational Methods}\label{methods}
All the systems used in this work for unbiased simulations were solutions of a single solute 
molecule with 1000 SPC/E\cite{SPCE_JPhysChem_Berendsen_1987} water molecules at water density 
\SI{0.997}{\gram\per\centi\meter\cubed}. The solutes studied were: an SPC/E water molecule, methane, 
n-octanol, benzene and the 20 proteinogenic amino acids. The amino acids were 
simulated in their standard physiological protonation state and with N-methylated and C-acetylated termini. 
The OPLS\cite{Kaminski2001} force field was used for methane and octanol.
AMBER03\cite{AMBERDuan2003} was used for the amino acids and benzene. The partial charges of benzene were 
calculated at the B3LYP/cc-PVTZ level using the ESP method\cite{MK2_MerzKollman_JCompChem_1990} and the 
polarizable continuum model \cite{PCM_Tomasi2005} was used to mimic the aqueous environment.
Water molecules were kept rigid using the SETTLE algorithm\cite{SETTLEmiyamoto1992settle}. For the rest of the 
solutes the bonds involving hydrogen were constrained with the P-LINKS algorithm\cite{PLINCS_Hess2008}. 
Lennard-Jones cross-term parameters were assigned using 
$\epsilon_{ij}=\left(\epsilon_{ii}\epsilon_{jj}\right)^{\left(1/2\right)}$ and 
$\sigma_{ij}=\frac{\sigma_{ii}+\sigma_{jj}}{2}$, except in the case of AMBER03 where 
$\sigma_{ij}=\left(\sigma_{ii}\sigma_{jj}\right)^{\left(1/2\right)}$ was used. 

The host-guest system studied by metadynamics simulation was obtained from the 
SAMPL5\cite{Yin2017} blind contest. The host-guest system studied has code name OAMe/OA-G2 
and 
the structure and topology files used were those provided for the contest. The force fields 
used were GAFF\cite{GAFFWang2004} and SPC/E\cite{SPCE_JPhysChem_Berendsen_1987}.

All molecular dynamics (MD) simulations were run with GROMACSv5.1.1\cite{GROMACS_Abraham2015} in the NVT ensemble using the stochastic velocity rescaling 
thermostat\cite{Bussi2007} at \SI{298}{\kelvin} and a relaxation time $\tau$=\SI{0.1}{\pico\second}. The equations of motion 
were integrated using the  leapfrog algorithm with a \SI{2}{\femto\second}  time step for a total time of 
\SI{10}{\nano\second}. In the case of the metadynamics simulation \SI{300}{\nano\second} 
although this is longer than necessary for convergence.
Periodic boundary conditions were used and long-range 
electrostatic interactions were calculated with the PME method\cite{PME1Darden1993,PME2Essmann1995}. Short 
range van der Waals interactions were truncated at \SI{10}{\angstrom}. 

The calculations of the fingerprint were done using a development version of PLUMED 2\cite{Tribello14}. 
The RDF is calculated using a kernel density estimation of the radial distribution function\cite{Piaggi2017,Piaggi2017b}. 
Which for a Gaussian kernel is:
\begin{equation}\label{mol}
\small
g_{i\text{w}}(r)=\frac{1}{4\pi \rho_{\text{w,loc}}r^2}
\sum_{j\in\text{w}}^{}\frac{1}{\sqrt{2\pi\sigma^2}}
e^{-\left(r-r_{j}\right)^2/\left(2\sigma^2\right)}
\end{equation}
where $r_j$ is the distance between the fingerprinted atom, $i$, and the $j$-th water molecule where $j$ runs 
over the set of water molecules. $\sigma$ is the 
Gaussian kernel bandwidth. Kernel density estimation ensures that $g_{i\text{w}}(r)$  is continuous and 
differentiable with respect to atomic positions for its use as a collective variable in 
enhanced sampling 
simulations. In addition, this decreases the noise when the statistics is poor. Nevertheless, a conventional 
RDF would give identical results. The value of $\sigma$ was \SI{0.05}{\angstrom} producing RDFs that are 
smooth but yet preserve all the relevant features. The fingerprint was integrated using the trapezoid rule.
The upper integration limit was chosen to be $r_{max}=$\SI{10}{\angstrom}.  \red{Equation 
(\ref{mol})  corresponds to the single configuration  $g_{i\text{w}}(r)$. To reduce noise,  
$g_{i\text{w}}(r)$ is averaged for its use in Equation 
(\ref{sij}).}

The local number density of water $\rho_\text{w,loc}$ is generally different from the bulk water density $\rho_\text{w}$.
This is a consequence of the excluded volume of the solute.
For big solutes such as the amino acids considered below, the deviation of $\rho_\text{w,loc}$ from $\rho_\text{w}$ can be very significant.
For this reason we have used the local density both in Equations \eqref{sij} and \eqref{mol}.
This choice ensures that the RDFs are all equivalently normalized regardless of the excluded volume of the solute.

Well tempered metadynamics (WTMetaD) simulations \cite{Barducci2008,Dama2014} were run on the host-guest 
system in its funnel variant\cite{Limongelli2013}. Funnel metadynamics adds a constant bias 
potential on the guest such 
that it remains in a funnel-shaped region with the conical part placed in the 
cavity of the host and the thin cylindrical region outside host. In this way 
the guest diffuses in a region of space where it can easily access the host and not diffuse 
through all space. The funnel has a length of \SI{23}{\angstrom} with the cone apex at 
\SI{15}{\angstrom} and cone angle of \SI{45}{\degree}. The funnel restrain was quadratic with 
a force constant of \SI{40}{\kilo\joule\per\angstrom\squared}.
The entropy loss due to this restrain is corrected analytically\cite{Limongelli2013,Allen2004} a posteriori using Equation 1 of the SI. 

WTMetaD was performed using two CVs: the inverse of the square root of host-guest contact map and a 
CV based on the local fingerprint \s that we shall refer to as \scv. 
Using the inverse of the square root of the contact map ensures that both states are sampled in a balanced fashion.
This compensates for the fact that a bound and unbound state have ranges of contact-map values that are very uneven. 
The chosen contacts are specified in PLUMED's input shown in the SI.
$S_{\text{CV}}$ is defined as the sum of the \s of several atoms of both the host and the guest. 
Only some solute atoms are included for the calculation of \scv in order to reduce their 
computational cost. The atoms used can found in Figure S6 of the Supporting Information (SI).
An additional simulation without biasing \scv was performed as a reference. 

The WTMetaD simulation was carried out using the same molecular dynamics parameters as the unbiased simulations. 
The Gaussians were deposited every \SI{1}{\pico\second} with an initial height of 
\SI{5}{\kilo\joule\per\mol}. The Gaussian $\sigma$s were 0.005 and 0.05$k_B$ for the contact 
map CV and the fingerprint CV. A bias factor of 24 was used. 
The free energy surfaces were reweighted by the method of Tiwary and Parrinello\cite{ReweightTiwary2015}. 
The statistical uncertainties are presented as the standard error of the mean calculated using 
block averages. 
Further details of the simulation can be found in the SI.

\section{Results and Discussion}
\subsection{Simple Solutes}\label{SimpleSolutes}

\begin{figure}
\centering
\includegraphics[width=0.9\columnwidth]{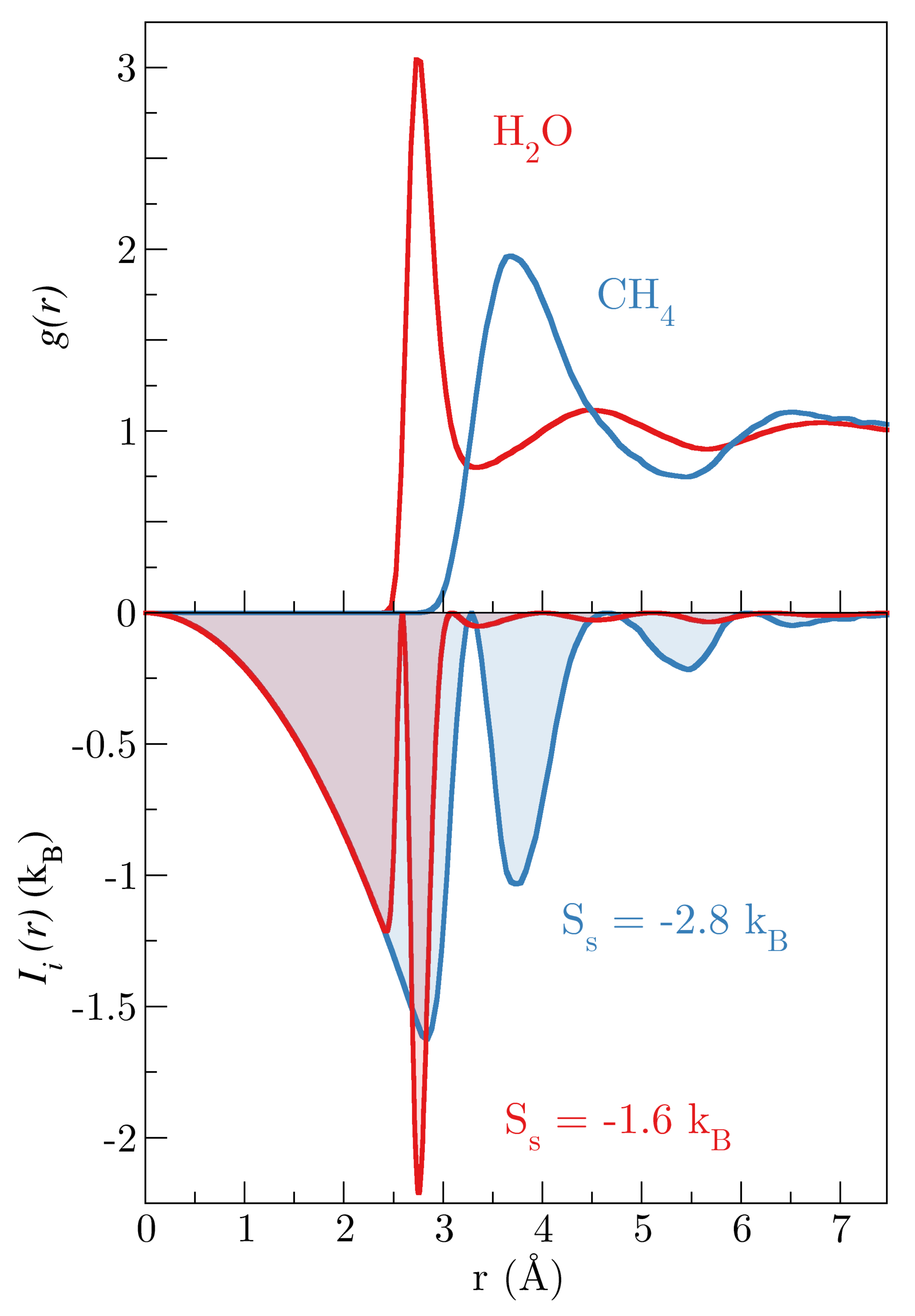}
\caption[]{Top: C-O radial distribution function for the aqueous methane 
simulation (blue) and O-O radial distribution function for a pure water simulation (red). Bottom: for the same 
pairs, the integrand, $I_i(r)$, of the 
fingerprint is plotted.}

\label{RDFInt}
\end{figure}

Water and methane are paradigmatic cases of hydrophilic and hydrophobic solutes. 
Thus, their local fingerprint values can be used as references. 
Water has an \s of $-1.57\pm0.01$ and methane of $-2.78\pm0.01$. 
Figure \ref{RDFInt} clarifies the physics behind these numbers. 
The top graph shows the radial distribution functions of the solutes and the bottom graph the 
integrand $I_i(r)$ of the fingerprint: 
 \begin{equation}
 \begin{split}
 I_i(r)=-2\pi 
\rho_{\text{W}}\left\{\right. &g_{i\text{w}}(r)\ln\left[g_{i\text{w}}(r)\right] \\
&-g_{i\text{w}}(r)+1\left.\right\}
 r^2
 \end{split}
 \end{equation}
The figure shows how \s varies 
with the radial structure of the solvent around the solute. In essence, \s becomes more 
negative 
the larger the deviation of the RDF from one.
The more the solvent is structured around the solute, the smaller \s.
Because of the $r^2$ factor, the structuring at larger distances is especially effective in 
decreasing \s.
At short distances, for $r$ less than a distance $r_c$ of the order of the molecular radius, 
$g_{i\text{w}}(r) \approx 0$ and this small $r$ 
region gives a contribution proportional to $r_c^3$.
This contribution to \s corresponds to the cavity formation entropy.
Methane has a lower \s than water for two reasons. 
First, it generates a larger cavity. 
Second, although its first hydration shell peak is less structured, it is wider, it is located 
at distances 
larger 
than the first hydration shell of water, and contains 4 times more water molecules.

We shall use the \s values for water and methane as representative of extreme hydrophilicity 
and 
hydrophobicity. It is therefore convenient to rescale the values of \s introducing an 
 index $h$ that is +1 for water and -1 for methane. Thus, in this scale the sign of $h$ 
determines whether the atom is hydrophobic or hydrophilic. 

\begin{figure}
\centering
 \includegraphics[width=0.9\columnwidth]{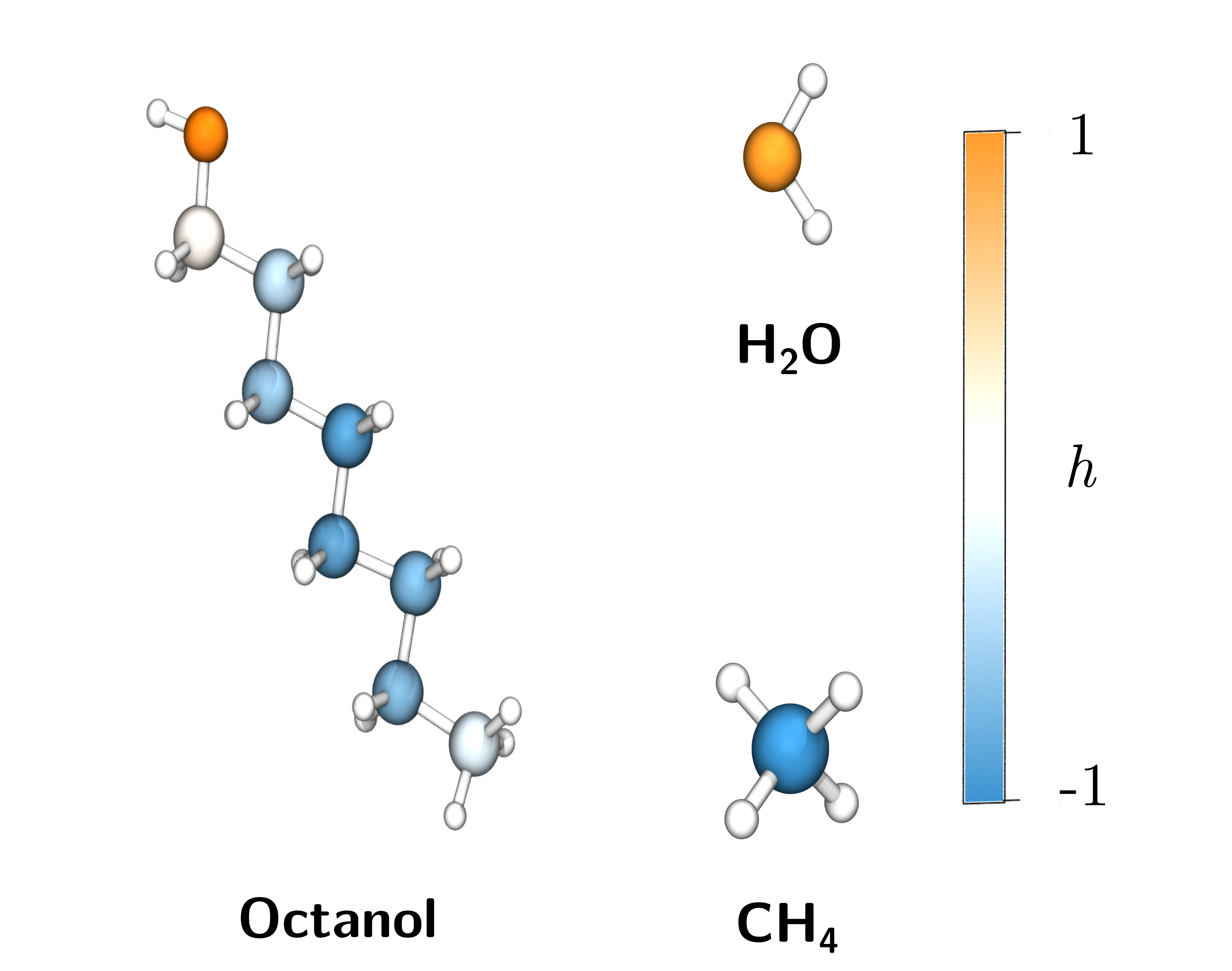}
\caption[]{Octanol, water and methane molecules with their heavy atoms colored according to 
the $h$ index.
The scale ranges from hydrophobic (blue), to intermediate (white) and to hydrophilic (orange).
}
\label{OCT}
\end{figure}

We now turn to discuss the properties of octanol chosen for its amphiphilic character.
Figure \ref{OCT}, shows octanol, water, and methane with their heavy atoms colored according 
to their $h$ values. 
The index clearly distinguishes 
between hydrophobic and hydrophilic atoms. The trend in $h$ values is in accordance with what 
could have been 
expected. The index can also deal with intermediate cases as the carbon 
atom attached to the alcohol group. This atom should be labeled as less hydrophobic than 
aliphatic 
carbons due to its partial positive charge generated by the electronegativity difference with 
the oxygen it is 
bonded to. 
Nevertheless, since the influence of a hydrophilic atom on the fingerprint of others is 
limited, the fingerprint is local with respect to the atoms of the molecule. 
The terminal \ce{CH3} has a lower $h$ than the \ce{CH2} carbons. 
This can be ascribed to the fact that the solvation shells of neighboring \ce{CH2} groups in 
the 
aliphatic chain overlap. This 
shifts the RDF first solvation peak to higher distances thus increasing $h$. Since \ce{CH3} 
has only one neighbor, this effect is less pronounced. Figure S3 of the SI 
illustrates this by analyzing 
the RDFs of primary, secondary, tertiary carbon atoms and methane.  Figure S4 of the SI 
includes the numeric 
values of the fingerprint of the atoms in octanol. 

An interesting case is that of ions, in which their classification into hydrophilic or 
hydrophobic could be misleading. 
The fingerprint \s for \ce{Na+} is -3.9\kb which would mistakenly classify it as more 
hydrophobic than methane. 
This is mostly due to the intensity of the first shell peak of the \ce{Na+-H2O} RDF which 
decreases strongly the value of \s because of 
the strong interaction with the ion (Figure S2 of the Supporting Information). 
In the classical electrochemistry or coordination chemistry notion of the hydrated 
ion\cite{chem_aq_ions_Richens}, we consider the ion and its first hydration shell as the 
solute. 
In this context we can consider the sodium cation as a buried atom and the first hydration 
shell atoms as the solvent exposed atoms in which to measure the fingerprint. 
This concept has been useful in the development of metal ion force 
fields\cite{JACS_ESM_1999,Angew_ESM_2010}. 
The first-shell water molecules have an \s of -0.9\kb which is more hydrophilic than bulk 
water. 
Therefore if we use the hydrated ion as the solute, we can conclude the \ce{Na+} hydrated ion 
is hydrophilic as expected.

\subsection{Amino acids}

\begin{figure}
\centering
\includegraphics[width=\columnwidth]{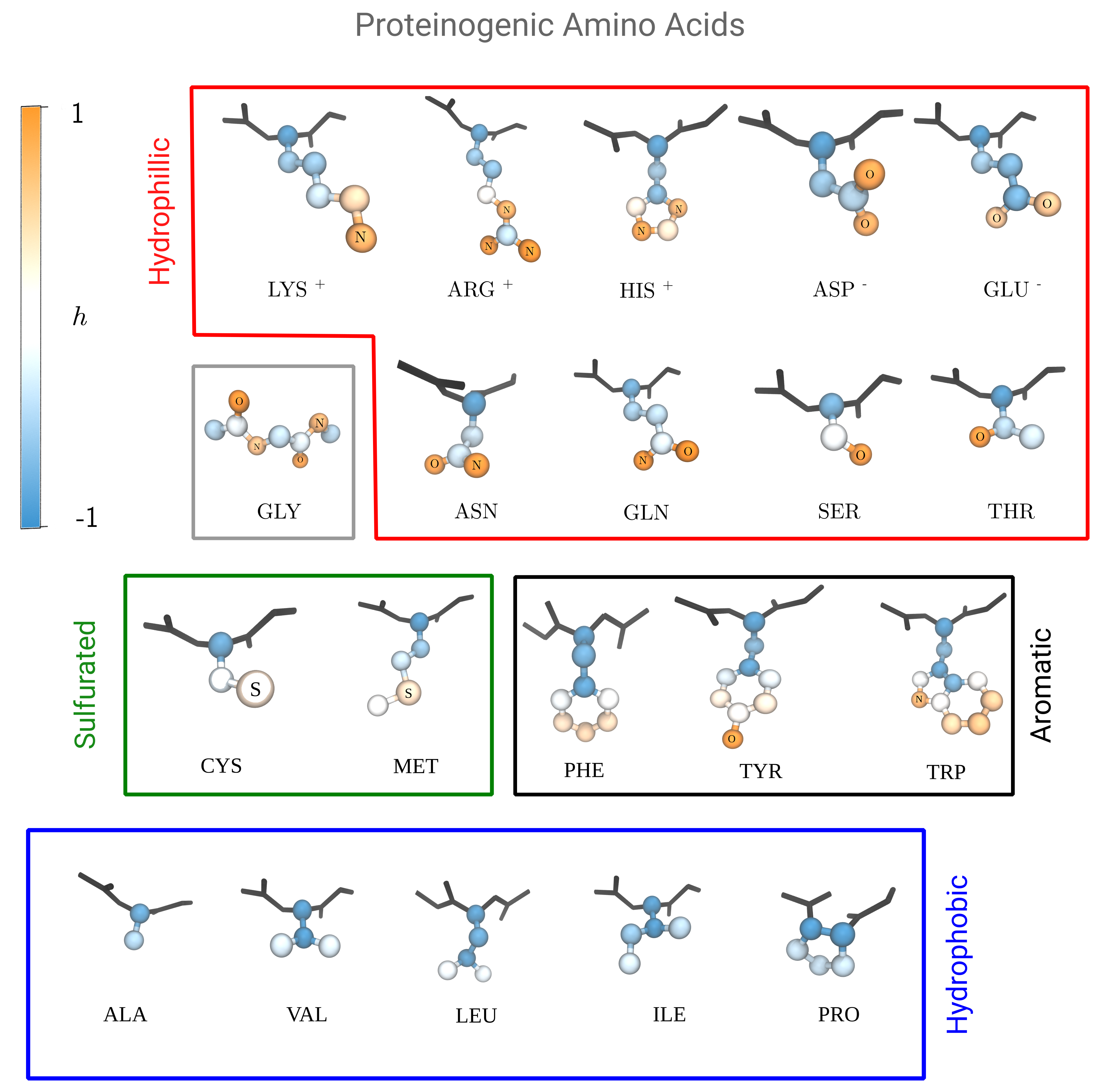}
\caption[]{Structures of the proteinogenic amino acids with their heavy 
atoms colored according to the $h$ index. The scale ranges from hydrophobic (blue), to 
intermediate (white) and to hydrophilic (orange). Unlabeled atoms are carbon. Hydrogen 
atoms are 
omitted. Since all backbone atoms have similar $h$ index, only side chain atoms are 
considered. Backbone atoms are visible for glycine (gray box). The boxes organize the amino 
acids by 
families: hydrophilic (red), glycine (gray), sulfur-containing (green), aromatic (black) and 
hydrophobic 
(blue). 
}
\label{AA}
\end{figure}

\begin{figure*}[!ht]
\centering
 \includegraphics[width=0.9\textwidth]{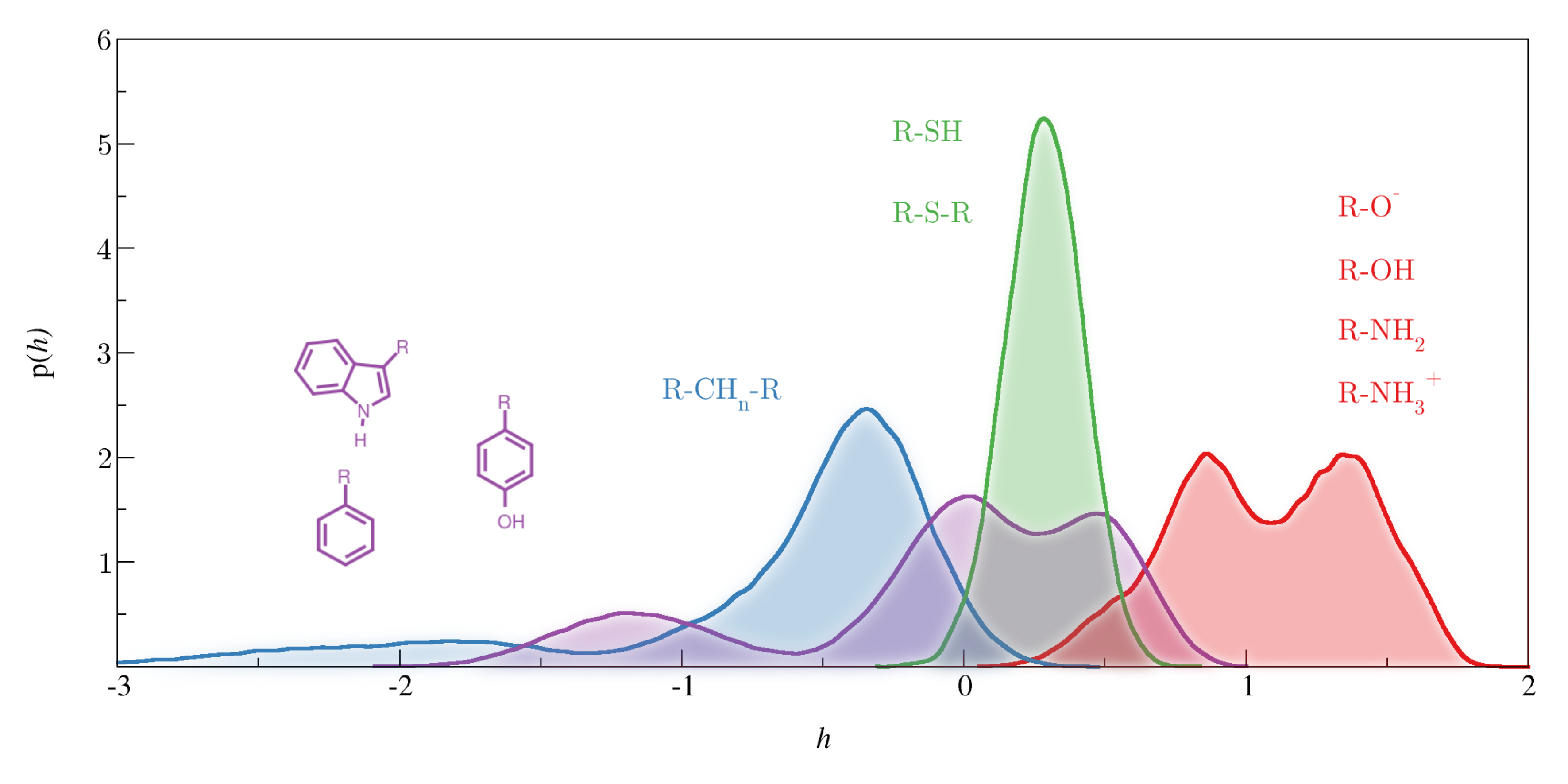}
\caption[]{ Probability densities of the hydrophobicity local fingerprint, $h$, of different groups 
of 
atoms in their respective simulations. The lines are the distributions of $h$ for:  C atoms of 
hydrophobic 
amino acids (blue), N and O atoms of hydrophilic amino acids (red), aromatic C atoms (purple) 
and  S atoms (green).}
\label{hist}
\end{figure*}

The local fingerprint for the heavy atoms of the 20 proteinogenic amino acids were computed, 
offering the 
possibility of testing our fingerprint on a wide range of chemical groups.
This is a first step for future use in the study of hydrophobic and hydrophilic interactions 
in proteins.
Figure \ref{AA} shows the different amino acid molecules with the heavy atoms in the side 
chains colored according to their $h$ value. The backbone atoms are shown only for glycine but 
a similar 
picture is obtained for the other amino acids. As in the case of octanol, $h$ assigns a 
hydrophobic value to aliphatic carbons and an hydrophilic value to polar N and O atoms of 
hydrophilic 
residues. All the heavy atoms of the backbone have $h$ values adequate to the hydrophilicity 
or hydrophobicity 
that chemical intuition suggests. 
A list of $h$ values can be found in Figure S5 of the SI.

While most of the $h$ values reflect the expected behavior, some apparently surprising values 
can be seen.
For instance the aromatic C are placed in the middle of the $h$ scale and thus they are 
classified as neither 
properly hydrophilic or hydrophobic. In reality this result is in line with the known 
solvation behavior of 
benzene which is much more soluble than its aliphatic counterpart cyclohexane.
The reasons for this effect have been discussed in the 
literature\cite{JPhysChem_McAuliffe_1966,JSolvChem_Cabani_1981,Ben-Amotz2016,Harris2016}.
As seen from the point of view of our fingerprint, this results from the fact that the other 
atoms in the ring exclude some of the solvation water leading to a reduction in the RDF peak 
height.
In order to confirm that this behavior is not an artifact of our force field, we have 
calculated \s using the 
benzene RDF kindly provided to us by Choudary et al obtained using ab initio 
MD.\cite{choudhary2016anisotropic} The ab initio value, \s $ = -1.9$\kb is very close to that 
of the AMBER 
force field. Here we did not scale the \s values since we do not have the ab initio reference 
point for 
methane. 

Another $h$ value that deserves some discussion is that of the sulfur atoms with a $h \sim 0$.
This can be linked to the fact that the electronegativity of sulfur is intermediate between 
carbon and oxygen, 
and to the ability of sulfur to accept weak H bonds\cite{Biswal2010,RaoMundlapati2015}.

Since we relate $h$ to the water solvation structure and the water structure around each atom and 
the conformation of the solute
can fluctuate as a function of time, we also looked at the distribution of this index. \red{We 
consider the $h$ value obtained from RDF averaged over a 400 \si{\pico\second} moving window 
to 
allow the 
local fingerprint to vary and study its distribution. The data obtained from all the amino acid 
simulations were put in a histogram in which we considered separately aliphatic C, aromatic C, S, 
and O and N of the side chains.}
The histograms are shown in Figure \ref{hist}.

In the histogram, the hydrophobic aliphatic C are clearly separated from the hydrophilic O and 
N of the side 
chains, proving the usefulness of the local fingerprint. 
As discussed previously, the distribution of the aromatic C and S are centered around $h \sim 
0$.
The distribution of the hydrophilic O and N of the side chains (shown in red in Figure 
\ref{hist}) presents two peaks and a shoulder.
The peak at $h \sim 0.8$ corresponds to all the hydrophilic O and N of charged amino acids 
with the exception of arginine, while the other peak at $h \sim 1.4$ corresponds to hydrophilic 
O and N of neutral amino acids and arginine.
Charged residues have a lower $h$ than neutral ones because they induce more structure in 
water.
Arginine is an exception to this rule due to its higher charge delocalization and therefore 
leads to a less 
well-defined solvation structure.
The shoulder at $0<h<0.6$ in the histogram of O and N of the side chains corresponds to 
glutamate since carboxylate oxygens have a very negative effective charge with respect to the 
rest of hydrophilic atoms.
The histogram of hydrophobic aliphatic C has two peaks.
The peak at $h \sim -0.4$ corresponds to \ce{CH3} carbon atoms, while the broad peak at $h 
\sim -1.75$ corresponds to \ce{CH2} and \ce{CH} carbon atoms.
This behavior has been discussed earlier in Section \ref{SimpleSolutes} for octanol. 
The histogram for aromatic C shows three peaks.
The two peaks around $h \sim 0.25$ correspond to the more solvent exposed aromatic C while the 
remaining peak centered at $h \sim -1.25$ corresponds to C closer to the C$\beta$. 

\subsection{Enhanced sampling simulations}\label{MetaD}
\begin{figure*}[!ht]
\centering
 \includegraphics[width=0.9\textwidth]{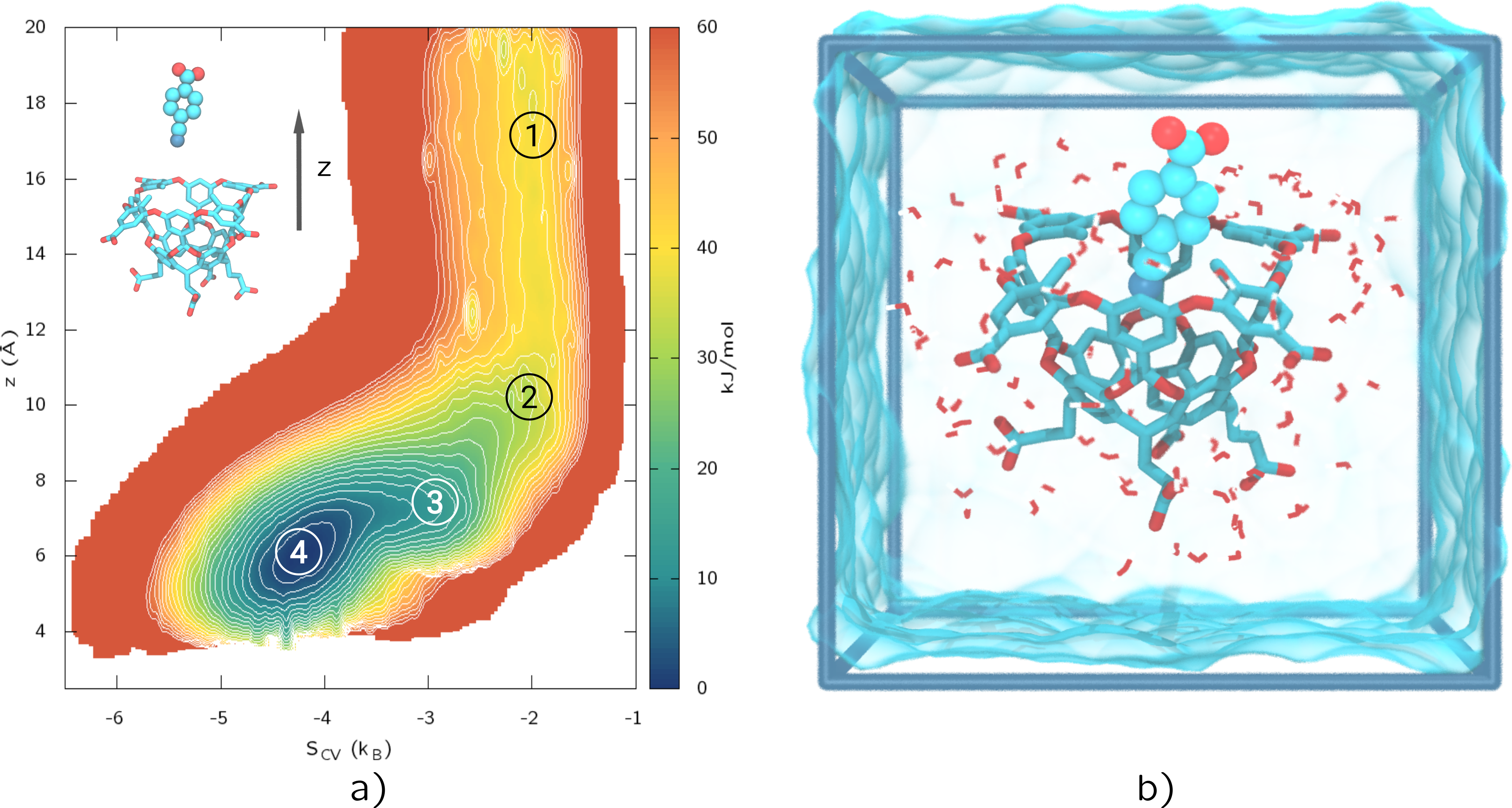}
\caption[]{ a) Reweighted free energy surface of the host-guest system as a function of the 
vertical distance between the centers of the guest and the bottom atoms of the host, $z$, and 
the fingerprint collective variable. b) Schematic (not to scale) representation of the 
host-guest system (OAMe-OAG2 in the SAMPL5\cite{Yin2017} contest). 
The solvation is representated by the surface and some of the water molecules are explicitly depicted.}
\label{FES}
\end{figure*}

In previous sections we used the local fingerprint to describe the hydrophobicity and hydrophilicity 
of different solutes.
In this section we will show that the local fingerprint can also be used as a collective variable to 
describe solvation in enhanced sampling simulations.
Figure \ref{FES}b shows the system chosen for the funnel WTMetaD simulations. 
It is a host-guest system consisting of a barrel-shaped host molecule and a ligand guest molecule that can fit in the cavity. 

As in many ligand-protein systems, desolvation is a key collective variable and a kinetic barrier to the binding if unbiased. 
If only the contact map is biased, the guest has to wait close to the entrance of the host 
until it desolvates and binding can happen (Figure S8 of the SI). 
As a consequence, the simulation lacks diffusion in CV space and the simulation's convergence is compromised. 
This has been observed for this system in previous metadynamics simulations by Bhakat et al.\cite{Bhakat2017a}.
Their solution was to add a static bias potential that desolvated the interior of the host during the metadynamics and then correct the free energy of binding with a disolvation free energy term obtained from a separate free energy perturbation simulation. 

Here we bias two CVs with WTMetaD: the inverse of the square root of host-guest contact map and a CV based on the fingerprint \s that we shall refer to as \scv. 
This results in convergence of the simulation and free diffusion of the system from bound to unbound 
(Figure S8). 
\scv acts as desolvation CV that allows the host and guest to desolvate during the binding 
process. 
From the simulation we calculate the free energy surface (FES) as a function of the vertical distance between the centers of the guest and the bottom atoms of the host, $z$, and \scv.
The FES is plotted in Figure \ref{FES}a.
In state 1) the guest is unbound and fully solvated. 
Along the diagonal path from 2) to 3), the guest is about \SI {6}{\angstrom} away from its 
bound position. The guest and host desolvate at the same time the guest enters the host. 
We can interpret this as the guest forcing water molecules out of the host-guest adduct as it 
is drawn by intermolecular forces into the opening of the barrel. 
Finally, from 3) to 4) there is a desolvation of the host and guest at nearly constant $z$. 
This is a situation in which the guest is at the host's doormat but requires a fluctuation of the solvent in order for there to be room in the host to enter. 
Our interpretation of the \scv as a desolvation CV is supported by the mirroring of the 
presented FES and an equivalent FES using the number of water molecules in 
the barrel instead of \scv.
This FES is shown in Figure S7 of the SI. 

Finally, the free energy of binding of the host to the guest is 
-28.1$\pm$0.8\si{\kilo\joule\per\mol}. This results from a projection of the FES onto 
$z$ and the entropy correction of the funnel. This result is close to the experimental 
value \SI{-21.6}{\kilo\joule\per\mol}\cite{Sullivan2017} and statistically identical to the 
value obtained by Bhakat et al.\cite{Bhakat2017a} using a different simulation protocol.

\section{Conclusions}

We have developed a local fingerprint for hydrophobicity and hydrophilicity.
The local fingerprint is inspired by the two body solute water contributions to the entropy which is 
a function of the RDF. 
In this context whether an atom is hydrophobic or hydrophilic is a consequence of the structure of water around it. 
This feature allows to understand how the character of a solute is modulated by its environment. 
We have also introduced an index of hydrophilicity $h$ that uses methane and water as representatives of hydrophobic and hydrophilic behavior. 
We show the usefulness of the fingerprint in enhanced sampling simulations by studying a host-guest system in which the fingerprint serves as a desolvation CV and allows for fast transition between the bound and unbound states. 
We expect that the fingerprint could also provide insight into more complex phenomena where hydrophobicity plays an important role, such as protein folding.

\begin{acknowledgments}

The authors thank Ashu Choudhary and Amalendu Chandra for kindly providing us with their ab initio benzene RDF.
The authors also thank Enrique S{\'a}nchez Marcos, Riccardo Capelli and Tarak Karmakar for helpful discussions. 
S.P-C. also acknowledges the Spanish Ministry of Education, Culture and Sports for the Ph.D. 
grant 
(FPU14/02100).
P.M.P and M.P acknowledge supported from the NCCR MARVEL funded by the Swiss National Science Foundation and from European Union Grant No. ERC-2014-AdG-670227/VARMET.
The computational time for this work was provided by the Swiss National Supercomputing Center (CSCS) under Project ID p503.
\end{acknowledgments}


\section*{Supporting information}

Supporting Information (PDF): Additional information including extra radial distribution functions, 
local fingerprint 
integrands, numeric values of the local fingerprint, free energy surfaces and metadynamics details 
not displayed in the text.

\end{document}